\documentclass[aps, twocolumn,superscriptaddress]{revtex4-2}
\usepackage{amsmath,amssymb,amsfonts,bbm,graphicx,times,psfrag}
\usepackage[pdftex]{color}
\usepackage{amsmath,bm}
\usepackage[colorlinks, linkcolor=blue, citecolor=blue, urlcolor=red, breaklinks=true]{hyperref}
\usepackage{microtype}

\usepackage{amsthm}
\usepackage{csquotes}
\usepackage{amsthm}\newcommand{\be} {\begin{equation}}
\newcommand{\ee} {\end{equation}}

\usepackage{color}

\begin{document}
\title{Strong mechanical squeezing in a microcavity with double quantum wells}
\author{Muhammad Asjad}
\affiliation{College of Computing and Mathematical Sciences, Department of Mathematics,Khalifa University, 127788, Abu Dhabi, United Arab Emirates}
\email{asjad\_qau@yahoo.com}
\author{Berihu Teklu}
\affiliation{College of Computing and Mathematical Sciences, Department of Mathematics, Khalifa University, 127788, Abu Dhabi, United Arab Emirates}
\affiliation{Center for Cyber-Physical Systems (C2PS), Khalifa University, 127788, Abu Dhabi, United Arab Emirates}
\email{berihu.gebrehiwot@ku.ac.ae}
\author{Hichem Eleuch}
\affiliation{Department of Applied Physics and Astronomy, University of Sharjah, Sharjah 27272, United Arab Emirates}
\affiliation{Institute for Quantum Science and Engineering, Texas A \& M University, College Station, TX 77843, USA}
\email{heleuch@sharjah.ac.ae}
\date{\today}
%%%%%%%%%%%%%%%%%%%%%%%%%%%%
\begin{abstract}
In a hybrid quantum system composed of two quantum wells placed inside a cavity with a moving end mirror pumped by  bichromatic coherent light, we address the formation of squeezed states of a mechanical resonator. The exciton mode and mechanical resonator interact indirectly via microcavity fields. Under the conditions of the generated coupling, we predict squeezing of the mechanical-mode beyond the resolved side-band regime with existing experimental parameters. Finally, we show that the robustness of this squeezing against thermal fluctuations is important for practical applications of such systems.\\

\end{abstract}
\maketitle
%%%%%%%%%%%%%%%%%%%%%%%%%%%%
\section{Introduction}
The generation and manipulation of optical squeezing has been of great theoretical interest since its development in the early days \cite{Caves,Yuen,Hollenhorst,Walls83,Walls}. These are quantum states of light characterized by reduced quantum noise in one observable at the expense of increased noise in the conjugate observable.  Methods for generating squeezed states generally couple the amplitude of a light beam to its phase. For example, it has been accomplished using nonlinear optical media \cite{Slusher}, where the index of refraction varies depending on the intensity of the light. Another promising possibility is optomechanical or “ponderomotive” squeezing \cite{Braginsky}, which involves coupling the quantum fluctuations in a light field to the mechanical motion of an oscillator. 
Optomechanical squeezing may provide the potential to be tunable to the mechanical frequencies of interest and to be low loss, however this technology has only recently been made possible by employing a variety of experimental platforms. 
 For example, a group of ultracold atoms in an optical cavity \cite{Brooks}, cavity polariton \cite{Bobrovska}, a micromechanical resonator coupled to a nanophotonic cavity \cite{Naeini} and a vibrating membrane in the center of an optical cavity \cite{Purdy}. The mechanical oscillators in all three situations have an efficient quality factor that oscillates at a sharp resonant frequency with little energy loss \cite{Keith,Fabre,Bose, Mancini, Vitali}. 

For a variety of possible applications, the interaction of laser light with a mechanical oscillator, such as a moving mirror, can be utilized to manipulate the quantum features of light.  It is well known that the optomechanical interaction is capable of generating squeezed states of light \cite{Purdy}. This is a resource for quantum enhanced sensing techniques \cite{Rossi, Marco20, Candeloro, Marco22, Asjad22}, in non-linear optics \cite{Purdy, Dong}, and as well as to examine quantum phenomena in macro systems \cite{Berihu, Purdy13, Berihu15, Albarelli, asjad,  Marinkovic, Delic, as}. The squeezing of the mechanical systems is of utmost importance in this situation. The generation of squeezed and entangled light fields by hybrid quantum-well optomechanical system may be utilized to carry out quantum information processing, understanding its dynamics in realistic conditions is crucial to the development of numerous quantum information technologies. Mechanical resonators have a tremendous potential for the construction of quantum-level on-chip devices \cite{Boulier}, quantum memory \cite{Mcg13, Set12, Set15}, generating mechanical quantum superposition states \cite{14}, optomechanically induced transpency \cite{21, 22} and employed as a means to transfer a quantum state from an optical cavity to a microwave cavity \cite{Tia12, Cle12}.  Quantum state transfer between separate parts is a crucial tool for developing quantum communications and information processing protocols \cite{Barzanjeh12, asjad15}. In this paper, motivated by some experimental implementations in hybrid quantum systems \cite{Fujisawa, Delbecq, Frey}, we analyze the possibility of generating mechanical squeezing in a hybrid system formed by an optomechanical containing two quantum wells and pumped by a bichromatic coherent light, which are more convenient to implement quantum effects \cite{Wiel, Hanson}. The significance of this work lies in the fact that it shows that quantum squeezing can be realized in hybrid quantum systems that are experimentally accessible and compatible with current technology. This opens up possibilities for implementing quantum effects and harnessing quantum resources in practical applications.  
The implications of these findings extend to the potential realization of photonic systems, where multiple quantum dots (QDs) are integrated with microcavities or waveguides. This suggests that such schemes could enable unprecedented levels of quantum control over collective degrees of freedom in nanoscale systems with mesoscopic numbers. Overall, the results presented in the paper pave the way for practical applications and advancements in quantum technology.

The structure of the paper as described is as follows: the scheme of the theoretical model, the Hamiltonian that describes the dynamical system, and the linearized quantum Langevin equations are presented in the next section. In Sec. \ref{sta}, the steady state minimum quadrature variance of the mechanical resonator is discussed. We analyze the influence of the different parameters on the squeezing of the mechanical mode. We summarize our results and conclude in the last section.

\begin{figure}[t] 
\begin{center}
\includegraphics[width=.5\textwidth]{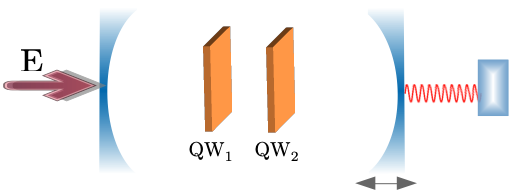}
\caption{Schematic representation of a nanoresonator formed by a microcavity, composed of a Fabry-Perot cavity defined by
 two sets of distributed Bragg reflectors (DBRs) and embedding two quantum wells (QWs) by placing
 at two different antinodes. The system is pumped by an external pump $E$ and has a damping
 rate $\protect\kappa$.
}
\label{fig1}
\end{center}
\end{figure}
\section{Model and Heisenberg-Langevin equations} \label{model}
In this  paper, we study a hybrid QW-microcavity system made up of two quantum wells placed at two different antinodes as shown in Fig. \ref{fig1}. The radiation pressure force in a microcavity and the exciton mode in the quantum well 
allow the mirror's optical and mechanical motion to be coupled. The exciton inside the QW in this model is seen as a quasi-particle that emerges from the interaction of one valence band hole with one
electron in the conduction band. We can treat exciton as a composed boson by assuming that the average distance between neighbouring excitons($\sim n_{\rm ex}^{-1/2}$ with $n_{\rm ex}$ being exciton concentration) is much larger than the  radius of the exciton. Since the density of the excitons is sufficiently low in the weak excitation regime, it is possible to neglect the contribution of the coupling of nearby excitons produced  by the Coulomb exciton-exciton interaction.  The interaction between nearby excitons, however, becomes strong and nonlinear in the moderate driving regime \cite{Cui00, Tas99, Cui98}, which results in intriguing features including squeezing and bistability \cite {Liu07, Ele10, Set11}. In the present work, we consider that the description of excitons as a composite of bosons is valid in the region of low density.

The Hamiltonian describes the dynamics of the coupled exciton-optomechanical system under the influence of a bichromatically driven optical cavity mode can be expressed as 

\begin{align}\label{ham}
H& =\omega _{a}a^{\dag }a+\omega _{\mathrm{m} }b^{\dag }b+\sum^2_{i=1} [ \omega _{\rm ex_i} c^{\dag }_i c_i  + ig_i(a^{\dag } c_i  - a c^{\dag }_i) ]\\ \notag
&-g_{0}a^{\dag }a (b+b^{\dag })  -i ( E^*(t) a -E(t) a^{\dag }). 
\end{align}%
Here the operators $a$, $b$, and $c_i$ (i=1,2) are annihilation operators for a photon in the microcavity, a phonon in the mechanical oscillator, and an exciton in the quantum well, respectively, $\omega _{a}$ and $\omega _{\mathrm{ex}}$ represent the microcavity and exciton resonance frequencies. The term $g_{0}$ represents the photon-optomechanical coupling, $\omega _{\mathrm{m}}$ is the resonance frequency of the mechanical resonator,  $g_i$ is the coupling strength of the cavity mode with \textit{ith} exciton mode, and  $E(t)=\varepsilon_+ e^{-i \omega_+ t}+\varepsilon_- e^{-i \omega_- t}$ is the two-tone laser field of amplitude  $\varepsilon _{\pm}=\sqrt{\kappa P_{\pm}/\hbar \omega _{\pm}}$ at frequency $\omega_{\pm}=\omega_{a}\pm\omega_{m}$ with $P_\pm$ and $\kappa$ represents the laser power and the cavity damping rate. The resulting steady-state amplitude of the intracavity shifts the equilibrium position of the mechanical mode via radiation pressure force.

In the Hamiltonian \eqref{ham}, the first three terms in the first line represent the free energy of the system while the last term describes the coupling of laser drive with the microcavity. In the second line, the first term describes the photon-phonon coupling and the fourth term in the first line represents the linear exciton-photon interaction.     
\begin{figure*}[!]
\begin{center}
\includegraphics[width=1\textwidth]{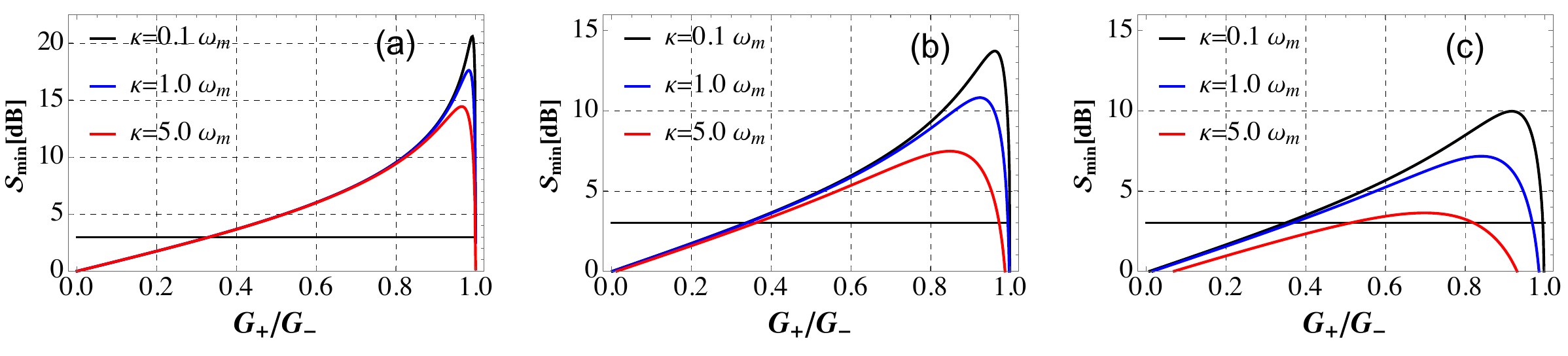}
\caption{ (a) The quadrature variance of mechanical resonator in dB with respect to vacuum level, i.e, $-10 {\rm Log}[\mathcal{S}_{\rm min}]$    $\mathcal{S}$ as function of ratio $G_+/G_-$ at $n_{\rm th}=0$, (b) $n_{\rm th}=10$, and (c) $n_{\rm th}=50$ for different values of the cavity decay rate $\kappa=0.1\omega_m$  (black curve),  $\kappa=1.0 \omega_m$  (blue curve) and $\kappa=5 \omega_m$ (red curve). The other parameters are $G_-=0.1\omega_m$, $g_1=g_2=2\omega_m$, $\Delta_{\rm ex_1}=-\Delta_{\rm ex_2}=\omega_m$, $\gamma=2\omega_m$, and $\gamma_m=10^{-5}\omega_m$. The horizontal solid black line corresponds to the 3dB limit.}
\label{fig2}
\end{center}
\end{figure*}

The system is affected by the environment's fluctuation and dissipation processes in addition to coherent dynamics. The following set of Heisenberg-Langevin equations are used to characterize the system dynamics as a function of dissipation and noise: 
\begin{eqnarray}
\dot {a}&=&-(\kappa+i\omega_{a}) a + \sum^2_{i=1} g_i c_i  + i g_{0} a (b+b^\dagger)+E(t) +\sqrt{2 \kappa}  a_{\rm{in}}  \notag,\\
\dot {b}&=&-(\gamma_{\rm m}+i\omega_m ) b+i g_{0}   a^\dagger a + \sqrt{2 \gamma_m} b_{\rm in}, \notag \\
\dot {c}_i &=&-(\gamma_i+i\omega_{\rm ex_i}) c_i -g_i a +\sqrt{2 \gamma_i}c_{\rm in_i},  \label{nql}
\end{eqnarray} 
where $\gamma_i$ is the excitonic dissipative coefficient related to the linear coupling to the excitonic bath describing the radiative and non-radiative dissipation, and $\gamma_{\rm m}$ is the mechanical mode damping rate. Here $a_{\mathrm{in}} $, $b_{\mathrm{in}}$, and $c_{\rm in_i}$ (for i=1,2) represent the input noise operators for the microcavity, exciton, and the mechanical modes, respectively. We assume that all the noise operators are Gaussian with the only terms with non-zero correlations
$\langle a_{\mathrm{in}}(\omega )a_{\mathrm{in}}^{\dag }(\omega ^{\prime })\rangle=2\pi \delta (\omega +\omega ^{\prime })$,  $\langle c_{\rm in_i}(\omega ) c_{\rm in_i}^{\dag}(\omega ^{\prime })\rangle =2\pi \delta (\omega +\omega ^{\prime })$,
$\langle b_{\rm{in}}(\omega )b_{\rm{in}}^{\dag }(\omega ^{\prime})\rangle  = 2\pi (n_{\rm{th}}+1)\delta (\omega +\omega ^{\prime })$, and $\langle b_{\rm{in}}^{\dag }(\omega )b_{\rm{in}}(\omega ^{\prime })\rangle  = 2\pi n_{\rm{th}}\delta (\omega +\omega ^{\prime })$, where $n_{\rm{th}}=[\exp (\hbar \omega _{\rm{m}}/k_{B}T)-1]^{-1}$ is the average amount of thermal excitation related to the mechanical resonator, where $k_{B}$ is the Boltzmann constant and T is the temperature of the mechanical bath. The nonlinear quantum Langevin equations in Eq.(\ref{nql}) can be linearized by considering the operators as the sum of the expectation plus quantum fluctuations. In order to linearize these equations, we can write the annihilation operators as $a=\langle a\rangle+\delta a$, $b=\langle b\rangle +\delta b$, and $c_i=\langle c_i\rangle+\delta c_i$. The equations of motion of the mean values of the operators can be evaluated by taking the quantum fluctuation average of the non-linear Heisenberg-Langevin equations Eq.(\ref{nql}), leading to
\begin{eqnarray}
\langle\dot {a} \rangle&=&-(\kappa+i\omega_{a}) \langle a\rangle + \sum^2_{i=1}g_i \langle c_i \rangle+ i 2g_{0}  \langle a \rangle \Re\langle b \rangle+E(t)  \notag,\\
\langle \dot {b} \rangle &=&-(\gamma_{\rm m}+i\omega_m )  \langle b \rangle+i g_{0}   |\langle a \rangle|^2,\notag \\
\langle\dot {c}_i \rangle &=&-(\gamma+i\omega_{\rm ex})  \langle c_i \rangle - g_i  \langle a \rangle.  \label{nem}
%-2i\alpha   \langle c \rangle   |\langle c \rangle|^2,  
\end{eqnarray} 
Here, we use the mean field approximation, i.e,$ \langle a c_i\rangle \approx \langle a\rangle \langle c_i\rangle$. The linearized Langevin equations of the fluctuation operators, in an interaction picture defined with respect to $\omega_a( a^\dagger a+c_1^\dagger c_1+c_2^\dagger c_2) + \omega_m b^\dagger b$ and applying the rotating wave approximation can be written as
\begin{align}
\delta \dot{a}&=-\kappa \delta a+\sum^2_{i=1} g_i \delta c_i+i(G_- \delta b+ G_+\delta b^{\dag })+\sqrt{2\kappa }a_{\rm{in}}, \nonumber\\
\delta \dot{b}&=-\gamma _{m}\delta b+i  (G_- \delta a + G_+ \delta a^{\dag })+\sqrt{2\gamma _{\rm{m}}}b_{\rm{in}},\nonumber \\
\delta \dot{c}_i&=-(\gamma_i+i\Delta_{\rm ex_i}) \delta c_i- g_i \delta a         %-i G_\alpha \delta c^{\dag }
+\sqrt{2\gamma_i }c_{\rm in_i}, 
\label{dc}
\end{align}
where $\Delta_{\rm ex_i}=\omega_{\rm ex_i}-\omega_a$ is the exciton-mechanical detuning,  $G_\pm=g_0 a_\pm$ are the dressed couplings between cavity mode and mechanical resonator with 
$$a_{\pm} =\varepsilon_\pm/\left(\kappa+i(\omega_a-\omega_\pm) + \sum^2_{i=1} g^{2}_i[\gamma_i+i \Delta_{\pm_i}]^{-1}\right)$$ and  $c_{\pm_i} = -i g_i a_{\pm}/(\gamma_i+i\Delta_{\pm_i})$,  where $\Delta_{\pm_i}=\omega_{\rm ex_i}-\omega_{\pm_i}$.
% +2\alpha (|c_+|^2+|c_{-}|^2). %and $G_\alpha=2\alpha c_+ c_-$ is the effective exciton-exciton scattering. 

\begin{figure*}[!]
\begin{center}
\includegraphics[width=1\textwidth]{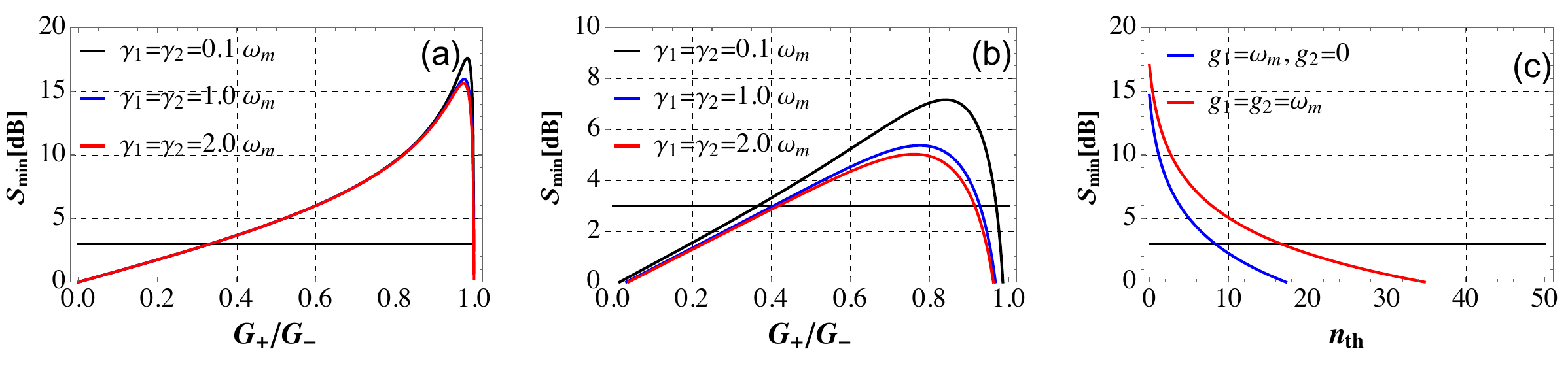}
\caption{ (a) The quadrature variance of mechanical resonator in dB with respect to vacuum level, i.e, $\mathcal{S}_{{\rm min}}[dB]=-10 {\rm Log}[\mathcal{S}_{\rm min}]$ as a function of ratio $G_+/G_-$ at $n_{\rm th}=0$ and (b) $n_{\rm th}=50$ for different values excitation cavity decay rate $\gamma_1=\gamma_2=0.1\omega_m$  (black curve),  $\gamma_1=\gamma_2=1.0 \omega_m$  (blue curve) and $\gamma_1=\gamma_2=2 \omega_m$ (red curve) . (c) $\mathcal{S}_{{\rm min}}[dB]$ vs thermal excitation $n_{th}$ for $g_1=\omega_m$, $g_2=0$ (blue curve) and  $g_1=g_2\approx \omega$ (red curve) at fixed value $G_+\approx 0.99\omega_m$ and $\kappa=\omega_m$. All other parameters are same as in Fig. \ref{fig2}.}
%The other parameters are $\Delta_{\rm ex_1}=-\Delta_{\rm ex_2}=\omega_m$, $\kappa=\omega_m$  and $\gamma_m=10^{-5}\omega_m$. The horizontal solid black line corresponds to the 3dB limit.}}}
\label{fig21}
\end{center}
\end{figure*}

\section{Stationary Mechanical squeezing} \label{sta}

It is more practical to investigate the mechanical mode's squeezing using the quadrature operators defined by, $\delta
X_a=(\delta a^{\dag }+\delta a)/\sqrt{2}$, $\delta Y_a=i(\delta a^{\dag}-\delta a)/\sqrt{2}$, $\delta
X_b=(\delta b^{\dag }+\delta b)/\sqrt{2}$, $\delta Y_b=i(\delta b^{\dag}-\delta b)/\sqrt{2}$, $\delta X_{c_i}=(\delta c^{\dag }_i+\delta c_i)/\sqrt{2}$,
and $\delta Y_{c_i}=i(\delta c^{\dag}_i-\delta c_i)/\sqrt{2}$. The fluctuation operators $\delta X_{j,\rm in}, \delta Y_{j,\rm in}$ ($j=a, b, c_i$).
Then, the set of linearized Heisenberg-Langevin equations for the quadrature operators in compact matrix form read
\begin{equation}\label{R}
\dot{u}= \mathbf{R} u+\mathbf{D}\mathbf{\eta},
\end{equation}%
where $u=\left(\delta X_{b}, \delta Y_{b},\delta X_a,\delta Y_a \delta X_{c_1},\delta Y_{c_1}, \delta X_{c_2},\delta Y_{c_2}\right)^{\textit{T}}$ is vector of quadrature of quantum fluctuation operators and $\mathbf{\eta}= \left(\delta X_{b,\rm in}, \delta Y_{b,\rm in}, \delta X_{a,\rm in},\delta Y_{a,\rm in}, \delta \delta X_{c_1,\rm in},\delta Y_{c_1,\rm in}, \delta X_{c_2,\rm in},\delta Y_{c_2,\rm in} \right)^{\textit{T}}$
is a vector of the input noise fluctuation operators, $\mathbf{D}= \text{diag}\left( \sqrt{2\gamma_m}, \sqrt{2\gamma_m}, \sqrt{2\kappa} ,\sqrt{2\kappa},  \sqrt{2\gamma_1}, \sqrt{2\gamma_1}, \sqrt{2\gamma_2}, \sqrt{2\gamma_2}\right)$ is a diagonal matrix and $\mathbf{R}$ is a drift matrix, and its elements can obtained from Eq.(\ref{dc}). Since the system is linearized and input states are Gaussian in nature with zero mean, the quantum fluctuations are thus a continuous three-mode Gaussian state, which can be completely characterized by the tripartite-covariance matrix $\mathbb{\mathcal{V}}$ of the exciton-optomechanical system defined as $\mathbb{\mathcal{V}}_{ij} =\langle\mathbf{u}_i \mathbf{u}_j+\mathbf{u}_j\mathbf{u}_i\rangle /2$. Then the covariance matrix  in steady state is determined by solving the Lyapunov equation $\mathbf{R}\mathbb{\mathcal{V}}+\mathbb{\mathcal{V}}\mathbf{R}^T+\mathcal{N}=0$, where $\mathcal{N}=\text{diag}\left( \gamma_m(2 n_{th} +1),\gamma_m(2 n_{\rm th} +1), \kappa, \kappa,  \gamma_1,\gamma_1, \gamma_2,\gamma_2  \right)$. 

To investigate the squeezing of the mechanical mode, we need to calculate the fluctuations of the mechanical quadrature $Q(\theta)=\delta X_{b} \cos(\theta) +\delta Y_{b} \sin(\theta)$. The state of the mechanical resonator is squeezed when the variance of $Q(\theta)$ is below the quantum noise level,\textit{i.e}, $\langle Q^2(\theta) \rangle<1/2$, and by minimizing the mean square value of $Q(\theta)$ with respect $\theta$, one gets 
\begin{align}
\mathcal{S}_{\rm min}&={\rm min }\langle Q^2(\theta) \rangle_{\theta} \notag\\
&=\mathbb{V}_{\rm p}+\mathbb{V}_{\rm q}-\sqrt{\left(\mathbb{V}_{\rm q}-\mathbb{V}_{\rm p} \right)^2 + 4 \mathbb{V}^2_{\rm qp} },
\end{align} 
where $\mathbb{V}_{\rm qp}$= $\{\mathcal{V}\}_{12}+\{\mathcal{V}\}_{21} $. Here $\mathbb{V}_{\rm p}$ and $\mathbb{V}_{\rm q}$ the first two diagonal elements of the covariance matrix $\mathcal{V}$  and represent amplitude and phase quadratures variances for the mechanical resonator, respectively.
For the effectiveness of this scheme, we make sure that the stability conditions are always satisfied. According to the Routh-Hurwitz criterion \cite{etb}, if all the eigenvalue of the drift matrix $\mathbf{R}$ in left-half of a complex plan (negative real parts)  (i.e., $|\mathbf{R}- \lambda \mathbf{I}| = 0), Re[\lambda]<0$, the system reaches its steady state and stable. Therefore we make sure that the stability conditions are all satisfied in the following section.
\begin{figure*}[!]
\begin{center}
\includegraphics[width=1\textwidth]{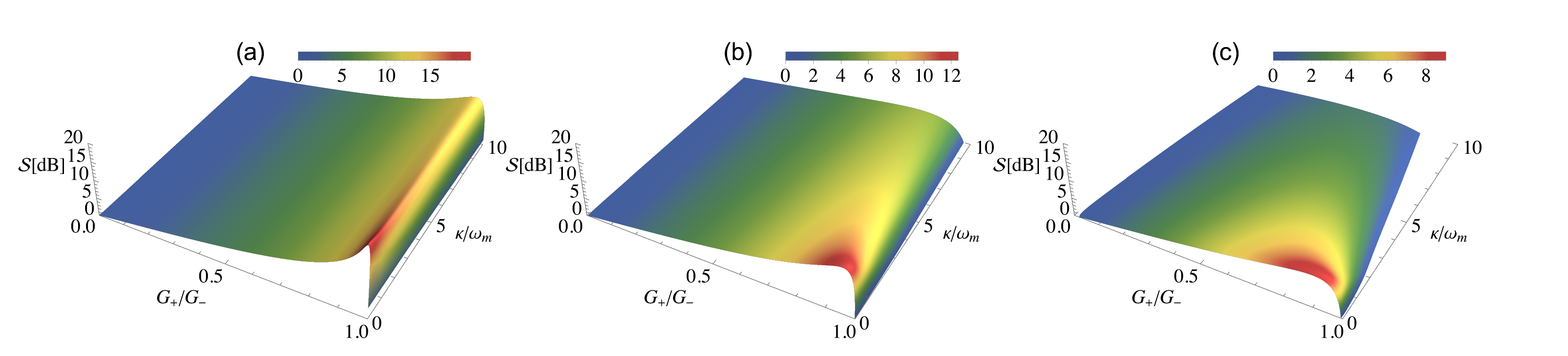}
\caption{ (a) The quadrature variance of mechanical resonator in dB with respect to vacuum level, i.e,  $\mathcal{S}_{{\rm min}}[dB]==-10 {\rm Log}[\mathcal{S}_{\rm min}]$ as function of ratio $G_+/G_-$ and cavity decay rate $\kappa/\omega_m$ at $n_{\rm th}=0$, (b) $n_{\rm th}=10$ and (c) $n_{\rm th}=50$. The other parameters are same as in { {Fig. \ref{fig2}.}}}
\label{fig3}
\end{center}
\end{figure*}

The evolution of the minimum quadrature variance $-10 {\rm Log}[\mathcal{S}_{min}]$ of the mechanical resonator in the steady state as function of ratio of the couplings $G_+/G_- $ for different values of cavity decay rate $\kappa=0.1\omega_m$(black curve), $1.0\omega_m$(blue curve), $5.0\omega_m$(red curve) and thermal bath phonon numbers $n_{\rm th}=0\, (a), \, 10 (b),\, 50 (c)$ as shown in Fig. (\ref{fig2}). For $\kappa=5\omega_m$, we notice that the mechanical resonator squeezing can break the 3 dB limit at large values of the  thermal bath phonon $n_{\rm th}=50$ as shown in Fig. \ref{fig2}(c) (red curve).  It is apparent from Fig. (\ref{fig2}) that the squeezing of the mechanical resonator decreases with increasing the decay rate of the cavity mode which means, the mechanical squeezing is very sensitive to the dissipation of the cavity mode. Furthermore, thermal excitation associated with the mechanical resonator influences the squeezing of the mechanical mode.  We find that mechanical squeezing still exists when the thermal occupation of phonon is $n_{\rm th}=50$ (in Fig. \ref{fig2}.(c)) and $n_{\rm th}=10$ (in Fig. \ref{fig2}.(b)), meaning that the mechanical squeezing is robust against the thermal environment  associated with mechanical mode.

{In Fig. \ref{fig21}, we depict the mechanical squeezing as a function of the ratio of two coupling strengths $G_+/G_- $ and the thermal excitation associated with mechanical mode $n_{th}$. For the environment thermal noise of $n_{th}=0$ (Fig. \ref{fig21}(a) ) and $n_{th}=50$ (Fig. \ref{fig21}(b)) respectively, we set the dissipation rates of the two exciton modes to $\gamma _1=\gamma _2=0.1\omega_m$ (black curve), $1.0\omega_m$ (blue curve) and $5.0 \omega_m$ (red  curve), the squeezing of the mechanical mode beyond 3 dB can be achieved even when
 $\gamma _1=\gamma _2=5.0\omega_m$ and $n_{th}=50$. The quadrature squeezing of mechanical resonator as a function of  the mean thermal excitation number of the phonon mode $n_{th}$ in the presence $g_1=g_2=\omega_m$ (red curve) and the absence of $g_1=\omega_m$, $g_2=0$  (blue curve) of second exciton mode $c_2$ is shown in Fig. \ref{fig21}(a). From  Fig. \ref{fig21}(c) we can see that in the presence of the second exciton mode $g_2=\omega_m$ (red curve), the squeezing of the mechanical mode can be achieved beyond 3 dB for relatively high values of the environment thermal noise $n_{th}$. We notice also in Fig3a. and (3b) that exists an optimal range of $G_+/G_- $ ( which  roughly between $0.4$ and $0.9$) to realize a maximum of squeezing for a moderate cavity dissipation ($\kappa=\omega_m$). 

In Fig. (\ref{fig3}), we plot the  squeezing of the mechanical resonator as a function of the dissipation rate of the optical mode $\kappa/\omega_m$ and the ratio of the couplings $G_+/G_- $ for different values of the mechanical excitation number $n_{\rm}=0$ Fig. \ref{fig3}(a), $n_{\rm th}=10$ Fig. \ref{fig3}(b) and $n_{\rm th}=50$ Fig. \ref{fig3}(c). It is clear from Fig. \ref{fig3} that the minimum quadrature variance is obtained by optimizing the coupling ratio $G_+/G_- $ and is not possible to squeeze the mechanical degree of freedom without  the blue-detuned laser  $(G_+ = 0)$. We notice that the mechanical squeezing can be obtained beyond the resolved side-band regime {($\kappa>\omega_m$) as shown in Fig. (\ref{fig3}).

\section{Conculusions}\label{conc}

In this paper we have addressed a scheme that demonstrates the potential for achieving squeezing beyond 3dB in a hybrid quantum well-optomechanical system 
by using a two tone drive and two exciton modes of different frequencies. We find that the mechanical squeezing obtained is robust against thermal noise associated with the mechanical resonator in the presence of two exciton modes. Furthermore, we have found that for achieving optimal 3dB squeezing performance of mechanical mode is only obtained for an appropriate value of the ratio of the coupling of the optical mode with mechanical resonator settings. The results presented in the paper contribute to the advancement of quantum technology by demonstrating the feasibility of implementing quantum effects in photonic systems. This paves the way for the development of practical applications that leverage quantum resources to achieve enhanced performance and functionality.

\section*{Acknowledgments}
\noindent M. Asjad has been supported by Khalifa University of Science and Technology under Award No. FSU-2023-014 and  8474000358 (FSU-2021-018). B. Teklu has been supported by Khalifa University of Science and Technology through the project no. C2PS-8474000137. 
%%%%%%%%%%%%%%%%%%%%%%%%%%%%%%%%%%%%%%%
%\bibliography{References}

\end{document}